\documentclass[12pt]{article}

\begin{document}

\begin{center}

{\Large {"Old" Conformal Bootstrap on AdS: 
\\
$O(N)$ symmetric scalar model}}\footnote{{\it{Int. J. Mod. Phys. A}}, {\bf 35}, 2050001 (2020)}

\vspace{1,5cm}

{Boris L. Altshuler}\footnote{E-mail addresses: baltshuler@yandex.ru $\,\,\,  \& \,\,\,$  altshul@lpi.ru}

\vspace{1cm}

{\it Theoretical Physics Department, P.N. Lebedev Physical
Institute, \\  53 Leninsky Prospect, Moscow, 119991, Russia}

\vspace{1,5cm}

\end{center}

{\bf Abstract:} Bootstrap equations for conformal correlators that mimic the early theory of conformal bootstrap are written down in frames of the AdS/CFT approach. The simplified version of these equations, that may be justified if Schwinger-Keldysh formalism is used in AdS/CFT instead of conventional Feynman-Witten diagrams technique, permits to calculate values of conformal dimensions in the $O(N)$ symmetric model with conformal or composite Hubbard-Stratonovich field.

\vspace{0,5cm}


\vspace{0,5cm}


\newpage

\tableofcontents

\newpage

\section{Introduction}

\quad Explanation of the values of fundamental constants now introduced {\it ad hoc} remains among the main challenges of the Standard Model and of theoretical physics as a whole. Steven Weinberg, when he was recently asked what single open question he would like to see answered in his lifetime, replied that it is only the mystery of the observed pattern of quarks and leptons masses \cite{Weinberg}. In higher dimensional theories spectra of physical particles are obtained as eigenvalues of equations for bulk fields, and it is possible to calculate the looked for constants with the choice of the bulk dynamics, including fields' bulk masses in units of the AdS curvature, and of boundary conditions. It is evident that there is great arbitrariness in this approach. Thus the theory determining the bulk masses would be a step in the desired direction.

"Old" conformal bootstrap considered in the AdS/CFT context allows, in principle, to calculate the values of conformal dimensions that is of bulk masses. To demonstrate this possibility is the goal of this paper.

The paper was inspired by the work \cite{Melon1} that Igor Klebanov has reported in the Lebedev Institute in 2017. In \cite{Melon1} (see also \cite{Melon2}), among other interesting things, the expressions for spectra of conformal dimensions in one-dimensional Sachdev-Ye-Kitaev and in $d$-dimensional field theory models were obtained from the "bootstrap" equations for Green function $G(X_{1}, X_{2})$ and vertex $\Gamma (X_{1}, X_{2}, X_{3})$ expressed symbolically in most general form as:

\begin{eqnarray}
\label{1}
G (X_{1}, X_{2}) = \int\int\, G(X_{1}, X)\,\Sigma (X, Y \, | \, G, \Gamma) \, G(Y, X_{2})\, dX\,dY \, ;  \nonumber
\\
\\
\Gamma (X_{1}, X_{2}, X_{3}) = M^{\triangle} (X_{1}, X_{2}, X_{3} \, | \, G, \Gamma),  \qquad  \qquad \nonumber
\end{eqnarray}
where $\Sigma$ is quantum self-energy and $M^{\triangle}$ is a triangle 3-gamma diagram - both constructed from the same $G$ and $\Gamma$ that stand on the LHS in (\ref{1}). 

The most characteristic feature of Eq-s (\ref{1}) is the absence there of any sign of the "bare" Lagrangian. In this sense it has something in common with Sakharov's zero Lagrangian approach \cite{Sakharov}. Schwinger-Dyson Eq-s of type (\ref{1}) with additional massless "bare" terms are used in theories with dynamical generation of mass (gap) such as superconductivity and models of spontaneous symmetry breaking.

In the "old" conformal bootstrap pioneered in \cite{old1}, \cite{old2} and developed in \cite{Parisi} - \cite{Dobrev2}, see e.g. \cite{Grensing} and references therein, the planar approximation $\Sigma \sim G^{2}$ is traditionally used in (\ref{1}) in self-consistent calculations of the anomalous conformal dimensions. In this case first line of (\ref{1}) takes a form (triple interaction is supposed, $g$ is the coupling constant): 

\begin{equation}
\label{2}
G (X_{1}, X_{2}) = g^{2} \, \int\int\, G(X_{1}, X)\,G(X, Y) \, G(X, Y) \, G(Y, X_{2})\, dX\,dY. 
\end{equation}
In some models the choice of only planar diagrams may be justified as "most divergent" ones or in frames of the $1/N$ expansion. However in his 1982 Nobel Lecture \cite{Wilson} Kenneth Wilson criticized this approximation as ungrounded. Nevertheless people use it, and we shall do the same in the present paper, having in mind that interesting results are the best justification of any postulate.  

"Old" conformal bootstrap  should not be confused with most popular nowadays "OPE conformal bootstrap" based on the demand of crossing symmetry of the operator product expansions - see primary papers \cite{Ferrara} - \cite{Mack} and modern reviews \cite{Rychkov}, \cite{Alday}.

To see the meaning of Eq. (\ref{2}) in the AdS/CFT context let us assume that $X_{1,2}$, $X$, $Y$ in (2) are the bulk coordinates in $AdS_{d + 1}$ and direct $X_{1, 2}$ to the horizon. Then LHS of (\ref{2}) becomes conformal correlator of the boundary conformal theory, whereas $G(X_{1}, X)$, $G(Y, X_{2})$ in the RHS become the corresponding bulk-to-boundary propagators. This procedure\footnote{I am grateful to Ruslan Metsaev for this observation.} is demonstrated in subsection 3.1 where the compact form of the exact equations of the self-energy AdS/CFT bootstrap are obtained.

Another background of the present paper are works \cite{Giombi1} where self-energy (bubble) Witten diagrams were calculated in case of triple bulk interactions, and \cite{Giombi2} where Witten diagrams with bulk Green functions replaced by the bulk harmonic functions were calculated.

In subsection 3.2 the simplified version of the self-energy bootstrap equation on AdS is presented where bulk Green functions in the corresponding Witten diagrams are replaced by their harmonic counterparts. Such a replacement was used earlier for scalar tadpoles in \cite{Mitra} - \cite{Alt2} and  for single-fermion loops in \cite{Allais}, \cite{Diaz2} in the context of the double-trace deformation. A possible justification of doing the same in the self-energy diagrams that are the product of two Green functions may be in the abandoning of the conventional Feynman-Witten diagram technique in favor of the Keldysh non-equlibrium Green functions ($NEGF$) technique, also called Schwinger-Keldysh, "in-in" or "closed time path" formalism, \cite{Keldysh}, see recent review \cite{Arseev}. $NEGF$ is widely used in the many particles theory, but it is also applied in a number of the early cosmology quantum problems with their strongly non-stationary de Sitter background, \cite{Barvinsky} - \cite{Akhmedov} and references therein. The possibility to apply $NEGF$ in AdS/CFT and to get in this way the "Keldysh component of self-energy" built from the two-point homogeneous solutions of dynamical equations is discussed in subsection 3.2.
 
In Sec. 4 calculation of values of conformal dimensions from the simplified bootstrap equations are performed in the $O(N)$ symmetric model of $N$ identical scalar fields interacting with the Hubbard-Stratonovich auxiliary scalar field. Since the self-energy bootstrap equations, in addition to the looked for fields' conformal dimensions, also contain unknown renormalized coupling constant,  the reduction of number of unknown variables is required to obtain a numerical results. In subsection 4.1 the option of conformally invariant Hubbard-Stratonovich field which conformal dimension is known is investigated; this option was inspired by \cite{Giombi1} where self-interacting conformal scalar field was studied in $AdS_{4}$. In subsection 4.2 the "extremal" relation imposed on values of conformal dimensions give the wishful reduction of unknowns; in this case we refer to \cite{Giombi2} where it was shown that calculation of Witten diagrams with extremal vertexes is essentially simplified.

3-point vertex of three interacting bulk scalar fields $\phi_{1}(Z)$, $\phi_{2}(Z)$, $\phi_{3}(Z)$ is called "extremal" when conformal dimensions of the fields obey the relation \cite{Hoker}:

\begin{equation}
\label{3}
\Delta_{\phi_{1}} = \Delta_{\phi_{2}} + \Delta_{\phi_{3}}.
\end{equation}
This extremal relation is valid in the elementary conformal theory when conformal operator $O_{\Delta_{\phi_{1}}}$ is the composite operator obtained by taking a product of two operators of dimensions $\Delta _{\phi_{2}}$ and $\Delta _{\phi_{3}}$. Following AdS/CFT approach we map the dual bulk scalar fields to each, including the composite one, conformal operator of the boundary conformal theory, while preserving the elementary relation (\ref{3}) between conformal dimensions when calculating Witten diagrams. This is consistent with the basic principles of the "old" conformal bootstrap that equates elementary conformal correlators to exact ones where quantum contributions are taken into account.

In Sec.2 familiar expressions used in the bulk of the paper are summed up. In Sec. 3 exact and simplified equations of the self-energy AdS/CFT bootstrap are obtained. Sec. 4 demonstrates the possibility of successful "hunting for numbers" of conformal dimensions in frames of the "old" conformal bootstrap in the AdS/CFT context. In Conclusion the possible directions of future work are outlined.

\section{Preliminaries}

\quad We work in $AdS_{d+1}$ in Poincare Euclidean coordinates $Z = \{z_{0}, {\vec z}\,\}$, where AdS curvature radius $R_{AdS}$ is put equal to one:

\begin{equation}
\label{4}
ds^{2} = \frac{dz_{0}^{2} + d {\vec z}\,^{2}}{z_{0}^{2}},
\end{equation}
and consider bulk scalar fields. Bulk field $\phi (X)$ of mass $m$ is dual to boundary conformal operator $O_{\Delta_{+}} (\vec x)$ or to its "shadow" operator $O_{\Delta_{-}}(\vec x)$ with scaling dimensions

\begin{equation}
\label{5}
\Delta_{\phi}^{IR} = \frac{d}{2} + \sqrt{\frac{d^{2}}{4} + m^{2}}, \, \, \, \,  \Delta_{\phi}^{UV} = d - \Delta_{\phi}^{IR}.
\end{equation}

We take normalization of the scalar field bulk-to-boundary operator $G^{\partial B}_{\Delta} (Z; \vec x)$ and of the corresponding conformal correlator like in \cite{Giombi1}, \cite{Giombi2}:

\begin{eqnarray}
\label{6}
G^{\partial B}_{\Delta} (Z; \vec x) = \lim_{\stackrel {x_{0} \to 0}{}} \left[\frac{G_{\Delta}^{BB} (Z, X)}{(x_{0})^{\Delta}}\right] = C_{\Delta}\, \left [\frac{z_{0}}{z_{0}^{2} + (\vec z - \vec x)^{2}}\right]^{\Delta}, \nonumber
\\
\\
 C_{\Delta} = \frac{\Gamma (\Delta)}{2\pi^{d/2}\Gamma \left(1 + \Delta - \frac{d}{2}\right)}, \qquad  \qquad \qquad  \qquad \nonumber
\end{eqnarray}
and:

\begin{equation}
\label{7}
<O_{\Delta}({\vec x}) O_{\Delta} ({\vec y})> = \lim_{\stackrel{x_{0} \to 0}{y_{0} \to 0}} \left[\frac{G_{\Delta}^{BB} (X, Y)}{(x_{0}\,y_{0})^{\Delta}}\right]= \frac{C_{\Delta}}{P_{xy}^{\Delta}}, \, \, \,  P_{xy} = |{\vec x} - {\vec y}|^{2}
\end{equation}

Bulk-to-bulk IR ($\Delta = \Delta^{IR} > d/2$, see (\ref{5})) scalar field Green function $G_{\Delta}^{BB} (X, Y)$ that is zero at infinity $x_{0}, y_{0} \to \infty$ has simple form in the boundary momentum space ${\vec p}$ ($p = |{\vec p}|$), and it possesses Kallen-Lehmann type spectral representation \cite{Penedones} - \cite{Bekaert}, \cite{Giombi1} (clear analysis of spectral representation for UV scalar field Green function is given in \cite{Giombi1}):

\begin{eqnarray}
\label{8}
G^{BB}_{\Delta} (X, Y) = \int_{-\infty}^{+\infty} \frac{\Omega_{\nu,0}(X,Y)\,d\nu}{[\nu^{2} + (\Delta - \frac{d}{2})^{2}]} = \qquad  \qquad  \qquad \qquad \nonumber
\\
\\
= (x_{0}y_{0})^{\frac{d}{2}}\,\int\frac{e^{-i{\vec p}\,({\vec x} - {\vec y})}\, d{\vec p}}{(2\pi)^{d}} \, [\theta (x_{0} - y_{0})\,K_{\Delta - \frac{d}{2}}(px_{0})\,I_{\Delta - \frac{d}{2}}(py_{0}) + (x_{0} \leftrightarrow y_{0})], \nonumber
\end{eqnarray}
$K$, $I$ are Bessel functions of the imaginary arguments. Nominator of the integrand in spectral representation in (\ref{8}) is scalar field Harmonic function that admits split representation and that is proportional to the difference (marked here with tilde) of IR and UV bulk Green functions:

\begin{equation}
\label{9}
\Omega_{\nu,0}(X,Y) = \frac{\nu^{2}}{\pi} \, \int \, G^{\partial B}_{\frac{d}{2} + i\nu}(X, {\vec x}_{a})\, G^{\partial B}_{\frac{d}{2} - i\nu}(Y, {\vec x}_{a})\, d^{d}{\vec x}_{a}  = \frac{i\nu}{2\pi}\,{\widetilde G}_{\frac{d}{2} + i\nu}, \qquad
\end{equation}

\begin{equation}
\label{10}
{\widetilde G}_{\Delta}(X, Y) = G_{\Delta}^{BB} - G_{d-\Delta}^{BB} = (d - 2\Delta)\, \int \, G^{\partial B}_{\Delta}(X, {\vec x}_{a})\, G^{\partial B}_{d-\Delta}(Y, {\vec x}_{a})\, d^{d}{\vec x}_{a};
\end{equation}
${\widetilde G}_{\Delta}(X, Y)$ is the difference of residues in poles $\nu = \pm i (\Delta - d/2)$ in the RHS of spectral representation (\ref{8}). Split representation in (\ref{9}), (\ref{10}) is immediately seen from the last line in (\ref{8}), in the boundary momentum space it is just a product of solutions of the homogeneous dynamical equation: ${\widetilde G} \sim K(px_{0})K(py_{0})$.

We shall also need expression for AdS/CFT tree 3-point vertex \cite{Freedman}, \cite{Penedones}, \cite{Paulos}, \cite{Giombi1}, \cite{Giombi2}:

\begin{eqnarray}
\label{11}
\Gamma_{\Delta_{1}, \Delta_{2}, \Delta_{3}} ({\vec x}_{1}, {\vec x}_{2}, {\vec x}_{3}) = 
\int \,G^{\partial B}_{\Delta_{1}} (X; {\vec x}_{1})\, G^{\partial B}_{\Delta_{2}} (X; {\vec x}_{2}) \, G^{\partial B}_{\Delta_{3}} (X; {\vec x}_{3}) \, dX =  \nonumber
\\  
\\
= \, \frac{B(\Delta_{1}, \Delta_{2}, \Delta_{3})}{P_{12}^{\delta_{12}}\,P_{13}^{\delta_{13}}\, P_{23}^{\delta_{23}}}, \qquad  \, \qquad \, \qquad  \, \qquad \, \qquad  \nonumber
\end{eqnarray}
where

\begin{equation}
\label{12}
\delta_{12} = \frac{\Delta_{1} + \Delta_{2} - \Delta_{3}}{2}; \,\,\, \delta_{13} = \frac{\Delta_{1} + \Delta_{3} - \Delta_{2}}{2}; \,\,\, \delta_{23} = \frac{\Delta_{2} + \Delta_{3} - \Delta_{1}}{2},
\end{equation}

\begin{equation}
\label{13}
B(\Delta_{1}, \Delta_{2}, \Delta_{3}) = \frac{\pi^{d/2}}{2}\, \left( \prod\limits_{i=1}^{3}\frac{C_{\Delta_{i}}}{\Gamma (\Delta_{i})}\right) \cdot \Gamma \left(\frac{\Sigma \Delta_{i} - d}{2}\right)\cdot \Gamma(\delta_{12})\, \Gamma(\delta_{13}) \, \Gamma (\delta_{23}).
\end{equation}

Also some well known \cite{Symanzik2}, \cite{Parisi}, \cite{Fradkin}, \cite{Giombi1}, \cite{Giombi2} conformal integrals will be used below:

\begin{equation}
\label{14}
\int \frac{d^{d}{\vec y}}{P_{1y}^{\Delta_{1}} \,P_{2y}^{\Delta_{2}}\, P_{3y}^{\Delta_{3}}} \stackrel {\Sigma \Delta_{i} = d} {=} \frac{A(\Delta_{1}, \Delta_{2}, \Delta_{3})}{P_{12}^{\frac{d}{2} - \Delta_{3}}\,P_{13}^{\frac{d}{2} - \Delta_{2}}\, P_{23}^{\frac{d}{2} - \Delta_{1}}},
\end{equation}
and

\begin{equation}
\label{15}
\int \frac{d^{d}{\vec y}}{P_{1y}^{\Delta_{1}} \,P_{2y}^{\Delta_{2}}} = \frac{A(\Delta_{1}, \Delta_{2}, d - \Delta_{1} - \Delta_{2})} {P_{12}^{\Delta_{1} + \Delta_{2} - \frac{d}{2}}},
\end{equation}
where

\begin{equation}
\label{16}
A(\Delta_{1}, \Delta_{2}, \Delta_{3}) = \frac{\pi^{d/2}\, \Gamma (\frac{d}{2} - \Delta_{1}) \, \Gamma (\frac{d}{2} - \Delta_{2}) \, \Gamma (\frac{d}{2} - \Delta_{3})}{\Gamma (\Delta_{1}) \, \Gamma(\Delta_{2}) \, \Gamma (\Delta_{3})}.
\end{equation}

We will also need integral (\ref{15}) when $\Delta_{1} + \Delta_{2} = d$, $\Delta_{1} \ne \Delta_{2} \ne d/2$ (the derivation of this formula is elementary in momentum space, see e.g. in \cite{Fradkin}):

\begin{equation}
\label{17}
\int \frac{d^{d}{\vec y}}{P_{1y}^{\Delta} \,P_{2y}^{d - \Delta}} = \frac{\pi^{d}\,\Gamma (\Delta - \frac{d}{2})\, \Gamma (\frac{d}{2} - \Delta)}{\Gamma (\Delta) \, \Gamma (d - \Delta)} \cdot \delta^{(d)} ({\vec x}_{1} - {\vec x}_{2}), \qquad \Delta \ne \frac{d}{2};
\end{equation}

and the divergent integral which is analyzed in detail in \cite{Giombi1}:

\begin{equation}
\label{18}
\int \frac{d^{d}{\vec y}}{P_{1y}^{\frac{d}{2}} \,P_{2y}^{\frac{d}{2}}} = \frac{A(\frac{d}{2}, \frac{d}{2}, 0)}{P_{12}^{\frac{d}{2}}} = \frac{\pi^{d/2} \, \Gamma (0)}{\Gamma(\frac{d}{2})} \cdot \frac{1}{P_{12}^{\frac{d}{2}}},
\end{equation}
(Eq-s (\ref{15}) and (\ref{16}) are used here in case $\Delta_{1} = \Delta_{2} = d/2$). $\Gamma (0)$ is written here symbolically, its possible different regularizations are discussed in \cite{Giombi1}.

\section{AdS/CFT bubble analogy of the self-energy "old" conformal bootstrap}

\subsection{Exact bootstrap equations.}

\qquad We consider here three bulk scalar fields $\phi_{i}(Z)$ ($i = 1, 2, 3$) with conformal dimensions $\Delta_{\phi_{i}}$ (\ref{5}) and triple bulk interaction:

\begin{equation}
\label{19}
L_{int} = g\,\phi_{1}(Z) \,\phi_{2}(Z) \,\phi_{3}(Z).
\end{equation}

As it was noted in the Introduction the AdS/CFT version of the "old" conformal bootstrap is obtained if coordinates $X_{1,2}$, $X$, $Y$ in general formula (\ref{2}) are considered as the bulk coordinates in $AdS_{d + 1}$ and $X_{1,2}$ are placed at the horizon with appropriate normalization like in (\ref{6}), (\ref{7}). This procedure transforms LHS of (\ref{2}) into boundary correlator (\ref{7}) whereas in the RHS of (\ref{2}) we'll have two bulk-to-boundary propagators (\ref{6}). Thus (\ref{2}) takes a form:

\begin{equation}
\label{20}
<O_{\Delta_{\phi_{1}}}({\vec x}_{1}) O_{\Delta_{\phi_{1}}} ({\vec x}_{2})> \, = \, {\cal M}^{{\rm 2pt \, bubble}}_{\Delta_{\phi_{1}}|\Delta_{\phi_{2}}\Delta_{\phi_{3}}}({\vec x}_{1}, {\vec x}_{2}),
\end{equation}
or

\begin{equation}
\label{21}
\frac{C_{\Delta_{\phi_{1}}}}{P_{12}^{\Delta_{\phi_{1}}}}  = g^{2} \int\int G^{\partial B}_{\Delta_{\phi_{1}}}(X ; {\vec x}_{1})\,G_{\Delta_{\phi_{2}}}(X, Y) \, G_{\Delta_{\phi_{3}}} (X, Y)\,G^{\partial B}_{\Delta_{\phi_{1}}}(Y ; {\vec x}_{2})\,dXdY.
\end{equation}
Index $\Delta_{\phi_{1}}|\Delta_{\phi_{2}}\Delta_{\phi_{3}}$ of ${\cal M}$ in (\ref{20}) indicates that bubble of the field $\phi_{1}$ is formed by the fields $\phi_{2, 3}$.

From spectral representations (\ref{8}) of Green functions of fields $\phi_{2,3}$ it follows that Witten diagram ${\cal M}^{{\rm 2pt \, bubble}}_{\Delta_{\phi_{1}}|\Delta_{\phi_{2}}\Delta_{\phi_{3}}}({\vec x}_{1}, {\vec x}_{2})$ in (\ref{20}), deciphered in the RHS of (\ref{21}), admits double integral spectral representation \cite{Giombi1}. Then, with account of (\ref{9}), bootstrap Eq. (\ref{21}) may be written as:

\begin{equation}
\label{22}
\frac{C_{\Delta_{\phi_{1}}}}{P_{12}^{\Delta_{\phi_{1}}}} = \int\int \, \frac{i\nu}{2\pi}\,\frac{i{\overline \nu}}{2\pi} \, \frac{d\nu\, d{\overline \nu} \, {\cal {\widetilde M}}^{{\rm 2pt \, bubble}}_{\Delta_{\phi_{1}}|\frac{d}{2} + i\nu,\frac{d}{2} + i {\overline \nu}}\,({\vec x}_{1}, {\vec x}_{2})}{[\nu^{2} + (\Delta_{\phi_{2}} - \frac{d}{2})^{2}] \, [{\overline \nu}^{2} + (\Delta_{\phi_{3}} - \frac{d}{2})^{2}]}.
\end{equation}
Here nominator ${\cal {\widetilde M}}$ in the integrand is built with the replacement of Green functions $G$ in the RHS of (\ref{21}) by their harmonic counterparts ${\widetilde G}$ (\ref{10}). We call this construction "harmonic bubble" and also mark it with tilde:

\begin{eqnarray}
\label{23}
{\cal {\widetilde M}}^{{\rm 2pt \, bubble}}_{\Delta_{\phi_{1}}|\Delta_{\phi_{2}}\Delta_{\phi_{3}}}\,({\vec x}_{1}, {\vec x}_{2}) =  \qquad  \qquad \qquad  \qquad \nonumber
\\
\\
= g^{2} \int\int G^{\partial B}_{\Delta_{\phi_{1}}}(X ; {\vec x}_{1})\,{\widetilde G}_{\Delta_{\phi_{2}}}(X, Y) \, {\widetilde G}_{\Delta_{\phi_{3}}} (X, Y)\,G^{\partial B}_{\Delta_{\phi_{1}}}(Y ; {\vec x}_{2})\,dXdY.  \nonumber
\end{eqnarray}
To use (\ref{23}) in spectral representation (\ref{22}) the substitutions $\Delta_{\phi_{2}} \to d/2 + i\nu$, $\Delta_{\phi_{3}} \to d/2 + i{\overline \nu}$ must be performed.

${\widetilde M}$ was calculated in \cite{Giombi1} (where it is designated as ${\cal F}^{\rm 2pt\,bubble}(\nu, {\overline \nu})$). The evident steps are as follows: (1) to use in (\ref{23}) split representations of ${\widetilde G}$ (\ref{10}); (2) to perform two bulk integrals that gives convolution of two vertexes (\ref{11}) over two boundary points ${\vec x}_{a}$, ${\vec x}_{b}$; (3) to perform familiar conformal integral (\ref{14}) over ${\vec x}_{b}$. Then (\ref{23}) comes to:

\begin{eqnarray}
\label{24}
{\cal {\widetilde M}}^{{\rm 2pt \, bubble}}_{\Delta_{\phi_{1}}|\Delta_{\phi_{2}}\Delta_{\phi_{3}}}\,({\vec x}_{1}, {\vec x}_{2}) = g^{2} \, (d - 2\Delta_{\phi_{2}}) (d - 2\Delta_{\phi_{3}})\,\frac{1}{P_{12}^{\Delta_{\phi} - \frac{d}{2}}}\, \int \frac{d{\vec x}_{a}}{P_{1a}^{\frac{d}{2}}P_{2a}^{\frac{d}{2}}}  \, \cdot \nonumber
\\  \nonumber
\\  \nonumber
\cdot \, B(\Delta_{\phi_{1}}, \Delta_{\phi_{2}}, \Delta_{\phi_{3}})\, B(\Delta_{\phi_{1}}, d - \Delta_{\phi_{2}}, d - \Delta_{\phi_{3}}) \, A(\delta_{12}, \delta_{13}, d - \Delta_{\phi_{1}}) = \quad \nonumber
\\
\\
= \frac{C_{\Delta_{\phi_{1}}}}{P_{12}^{\Delta_{\phi_{1}}}} \, \cdot \, \frac{g_{R}^{2}}{F(\Delta_{\phi_{1}})} \, \cdot \,  {\cal {\bf R}}(\Delta_{\phi_{1}}, \Delta_{\phi_{2}}, \Delta_{\phi_{3}}),  \qquad  \qquad \qquad \nonumber
\end{eqnarray}
divergent conformal integral (\ref{18}) is absorbed here, together with some coefficients, in the "bare" $g^{2}$ defining the renormalized coupling constant as:

\begin{equation}
\label{25}
g_{R}^{2} = g^{2}\,\frac{P_{12}^{\frac{d}{2}}}{32\pi^{d}}\,\int \frac{d{\vec x}_{a}}{P_{1a}^{\frac{d}{2}}P_{2a}^{\frac{d}{2}}}.
\end{equation}
Coefficient ${\cal {\bf R}}(\Delta_{\phi_{1}}, \Delta_{\phi_{2}}, \Delta_{\phi_{3}})$ is symmetric in three its arguments and is equal to:

\begin{eqnarray}
\label{26}
{\cal {\bf R}}(\Delta_{\phi_{1}}, \Delta_{\phi_{2}}, \Delta_{\phi_{3}}) = \Gamma\left(\frac{{\Sigma_{i}\Delta_{\phi_{i}}} - d}{2}\right) \, \Gamma\left(\frac{2d - {\Sigma_{i}\Delta_{\phi_{i}}}}{2}\right) \,  \cdot \nonumber
\\
\\
\cdot \, \frac{\Gamma(\delta_{12}) \, \Gamma(\delta_{13}) \, \Gamma(\delta_{23}) \,\Gamma\left(\frac{d}{2} - \delta_{12}\right) \, \Gamma\left(\frac{d}{2} - \delta_{13}\right) \, \Gamma\left(\frac{d}{2} - \delta_{23}\right)}{\Pi_{i=1}^{3}\left[\Gamma\left(\frac{d}{2} - \Delta_{\phi_{i}}\right) \, \Gamma\left(1 + \Delta_{\phi_{i}} - \frac{d}{2}\right)\right]}, \nonumber
\end{eqnarray}
and we introduced for brevity:

\begin{eqnarray}
\label{27}
F(\Delta) = \frac{\Gamma(\Delta) \, \Gamma(d - \Delta)}{\Gamma\left(\Delta - \frac{d}{2}\right) \, \Gamma\left(\frac{d}{2} - \Delta\right)}; \qquad  \qquad  \qquad   \qquad  \qquad  \nonumber
\\
\\
F_{d=1}(\Delta) = \frac{(\Delta - 1/2) \, \cos\pi\Delta}{\sin\pi\Delta}; \qquad F_{d=4} = (\Delta - 1)(\Delta - 2)^{2}(\Delta - 3);  \quad  \nonumber
\\  \nonumber
\\ \nonumber
F_{d=2}(\Delta) = - (\Delta - 1)^{2}; \,\,F_{d=3}(\Delta) =  - \frac{(\Delta - 1) (\Delta - 3/2) (\Delta - 2)\, \cos\pi\Delta}{\sin\pi\Delta}.  \nonumber
\end{eqnarray}

Third line in (\ref{24}) and expressions (\ref{25}) - (\ref{27}) are received with account of formulas for $A$, $B$, $\delta_{ij}$, and also $C_{\Delta}$ hidden in $B$, given in (\ref{16}), (\ref{13}), (\ref{12}), (\ref{6}) correspondingly.

We singled out in (\ref{24}) in front of the final expression for ${\widetilde M}$ the boundary conformal correlator (\ref{7}) that will reduce with the same correlator in the LHS of bootstrap equation (\ref{22}). 

Final expression for ${\cal {\widetilde M}}^{{\rm 2pt \, bubble}}_{\Delta_{\phi_{1}}|\Delta_{\phi_{2}}\Delta_{\phi_{3}}}\,({\vec x}_{1}, {\vec x}_{2})$ in the last line in (\ref{24}) (see ${\cal {\bf R}}$ in (\ref{26}), $\delta_{ij}$ in (\ref{12})) may be used in the integrand of exact bootstrap Eq. (\ref{22}) after changing the variables $\Delta_{\phi_{2}} \to d/2 + i\nu$, $\Delta_{\phi_{3}} \to d/2 + i{\overline \nu}$. It is seen from (\ref{26}) that exponential decay of Gamma functions at large imaginary arguments results in the exponential convergence of integrals in (\ref{22}) over $\nu$ for ${\overline \nu} = const$ and vise versa. Whereas in directions $\nu + {\overline \nu} \to \pm \infty$ ($\nu - {\overline \nu} = const$) or $\nu - {\overline \nu} \to \pm \infty$ ($\nu + {\overline \nu} = const$) when ${\cal {\bf R}}(\Delta_{\phi_{1}}, d/2 + i\nu, d/2 + i{\overline \nu}) \to const$ the double-integral in (\ref{22}) is divergent; thus bubble needs regularization. We will not go this way considered in \cite{Giombi1} but will study in the next Section the simplified UV-finite toy version of the bootstrap equations (\ref{22}).

Two more bootstrap equations are obtained from (\ref{22}), with account of (\ref{24}), (\ref{26}), (\ref{27}) by the permutation of fields. Thus we have three spectral equations for four unknown variables: $\Delta_{\phi_{1}}$, $\Delta_{\phi_{2}}$, $\Delta_{\phi_{3}}$ and $g_{R}^{2}$.

The most consistent way to get the missing fourth equation for coupling constant would be to require the validity of the vertex bootstrap equation in (\ref{1}) where three bulk variables $X_{i}$ are placed at the horizon. In AdS/CFT this means that vertex (\ref{11}) is equated to the one-loop triangle Witten diagram. This bootstrap equation is easy to put down, but it is difficult to work it out.

Another option is to reduce the number of unknown variables. We'll consider below in Sec. 4 two possibilities: 

1) to fix conformal dimension of one of the fields; the $O(N)$ symmetric model of $N$ scalar fields universally interacting with the Hubbard-Stratonovich conformally invariant field which conformal dimension is known will be considered in subsection 4.1;

2) to impose the "extremal" relation between the conformal dimensions of fields, see subsection 4.2.

\subsection{Simplified bootstrap equations: from Feynman-Witten to Schwinger-Keldysh?}

\qquad However, even reducing the number of unknown variables, it is not easy to solve the bootstrap Eq-s (\ref{22}) and those ones received from (\ref{22}) by permutation of fields. It would be interesting to find solution to this problem. In order to demonstrate that bootstrap is a working approach capable to predict the numerical values of conformal dimensions let us look at the simplified version of Eq-s (\ref{22}).  

Many poles structure of integrand in the RHS of (\ref{22}), including poles of the product of Gamma functions (\ref{26}), was investigated in \cite{Giombi1}. We will simplify bootstrap equations as much as possible, leaving in spectral integrals in (\ref{22}) only the residues at four poles corresponding to zeros of square brackets in the denominator in (\ref{22}). This actually means that in the bubble Witten diagram Feynman Green functions $G$ (\ref{8}) are replaced by their harmonic counterparts ${\widetilde G}$ (\ref{10}). Then instead of (\ref{22}) with conventional bubble ${\cal M}$ in the RHS we have bootstrap equation with harmonic bubble ${\cal {\widetilde M}}$ (\ref{23}) in the RHS:

\begin{equation}
\label{28}
\frac{C_{\Delta_{\phi_{1}}}}{P_{12}^{\Delta_{\phi_{1}}}} = {\cal {\widetilde M}}^{{\rm 2pt \, bubble}}_{\Delta_{\phi_{1}}|\Delta_{\phi_{2}},\Delta_{\phi_{3}}}\,({\vec x}_{1}, {\vec x}_{2}).
\end{equation}

Explicit expression (\ref{24}) for ${\widetilde M}$ permits to find values of conformal dimensions in Sec. 4 below. But what can justify such a fundamental departure from Witten's diagram technique?

The replacement of single Green functions (\ref{8}) by the "harmonic" difference (\ref{10}) of the corresponding IR and UV Green functions was used earlier in \cite{Mitra} - \cite{Diaz2} in calculations of scalar tadpoles and of single-fermion loops in context of the double-trace deformation. The possible justification of doing the same in bubble diagrams formed by the product of two Green functions may be in the application in AdS/CFT of Keldysh $NEGF$ (non-equilibrium Green functions) technique, also called Schwinger-Keldysh, "in-in" or "closed time path" formalism, that is used as a tool to cope with many-particle non-stationary problems, see pioneer papers  \cite{Keldysh} and recent review \cite{Arseev}. This formalism was also applied in a number of the early cosmology quantum problems with their strongly non-stationary de Sitter background, \cite{Barvinsky} - \cite{Akhmedov} and references therein.

In Witten diagrams built on the AdS background (\ref{4}) role of "time" plays coordinate $z_{0}$ and Green functions that are the building blocks of Witten diagrams are the exact copies of Feynman casual Green functions where ordinary time is replaced by $z_{0}$ (see e.g. in (\ref{8})). Since background (\ref{4}) is essentially "non-stationary", that is it strongly depends on "time" $z_{0}$, the question arises: why in the AdS/CFT calculations the traditional Feynman-Witten diagrams technique is used, and not $NEGF$ closed "time" path formalism? In this formalism the number of fields is doubled (according to doubling of time directions to forward and backward ones) and Green functions become the $2 \times 2$ matrices which diagonal elements are Feynman casual and anti-casual Green functions whereas non-diagonal terms are homogeneous solutions of dynamical equations.

Thus if we {\it postulate} that the endpoints of the bubble in bootstrap Eq. (\ref{21}) are located on different branches of the closed "time" loop, then this bubble will be the so-called Keldysh component of self-energy constructed only from the homogeneous  off-diagonal elements of the matrices of two bubble Green functions. In other words this means replacing of $T$-product of currents $T(j(X)j(Y))$ ($j(X) = \phi_{2}(X)\phi_{3}(X)$ for example) by the product of currents $j(X)j(Y)$. 

Of course, what was said above in support of the simple bootstrap equation (\ref{28}) is just a suggestive considerations. In any case the possibility to apply $NEGF$ technique in the AdS/CFT business deserves special attention by itself to our mind. And it is of interest to look at some consequences of simple bootstrap equation (\ref{28}). This is done in the next Section.

\section{Hunting for numbers}

\subsection{$O(N)$ symmetric model with conformal scalar.}

\qquad The $O(N)$ symmetric model of $N$ scalar fields $\psi_{k}$ with quartic interaction term $\sim (\Sigma_{k}\psi_{k}^{2})^{2}$ may be always reduced to triple interaction

\begin{equation}
\label{29}
L_{int} = g \, \sigma(Z)\,\Sigma_{k}\psi_{k}^{2}(Z)
\end{equation}
with the introduction of the auxiliary Hubbard-Stratonovich field $\sigma(Z)$. Let us consider theory of $N + 1$ scalar fields with interaction (\ref{29}) where $N$ fields $\psi_{k}$ have one and the same conformal dimension $\Delta_{\psi}$ and field $\sigma (Z)$ is conformally invariant in $AdS_{d + 1}$ that is its conformal dimension is known:

\begin{equation}
\label{30}
\Delta^{\rm conf}_{\sigma} = \frac{d}{2} \pm \frac{1}{2}.
\end{equation}

Then there are $N$ identical simplified bootstrap Eq-s (\ref{28}) written for fields $\psi_{k}$ (it is necessary to put in (\ref{28}) $\Delta_{\phi_{1}} = \Delta_{\phi_{2}} = \Delta_{\psi}$; $\Delta_{\phi_{3}} = \Delta_{\sigma}$), and one more bootstrap Eq. for field $\sigma$ when RHS of (\ref{28}) must be multiplied by $N$ since in (\ref{29}) $\sigma(Z)$ interacts with every of fields $\psi_{k}$ (in this case we must put in (\ref{28}) $\Delta_{\phi_{1}} = \Delta_{\sigma}$ and $\Delta_{\phi_{2}} = \Delta_{\phi_{3}} = \Delta_{\psi}$).

After these substitutions and with account of (\ref{24}), (\ref{26}), (\ref{27}) bootstrap Eq-s (\ref{28}) come to the system of two equations that excluding $g_{R}^{2}$ gives following spectral Eq. for $\Delta_{\psi}$:

\begin{equation}
\label{31}
N \, F(\Delta_{\psi}) = F(\Delta_{\sigma}).
\end{equation}

For $d = 4$ with account of expression (\ref{27}) for $F_{d = 4}(\Delta_{\psi})$ and value (\ref{30}) of $\Delta_{\sigma}$ when $F_{d = 4}(\Delta^{\rm conf}_{\sigma}) = - 3/16$ spectral equation (\ref{31}) comes to simple form:

\begin{equation}
\label{32}
N \, (\Delta_{\psi} - 1) (\Delta_{\psi} - 2)^{2} (\Delta_{\psi} - 3) = - \frac{3}{16}.
\end{equation}

The nontrivial positive roots of (\ref{32}) for three values of $N$ are: 

\begin{eqnarray}
\label{33}
N = 1: \Delta_{\psi} = 2 - \sqrt{3}/2 = 1.134; \nonumber
\\
N = 2: \Delta_{\psi} = 1.052; \,  \,  \,  1.680; \, \qquad
\\
N = 3: \Delta_{\psi} = 1.034; \,  \,  \,  1.741.  \, \qquad \nonumber
\end{eqnarray}
We did not show here the "shadow" conformal dimensions $4 - \Delta_{\psi}$ that also are the roots of (\ref{32}). All these values of conformal dimensions satisfy the unitarity bound $0 < |\Delta - d/2| < 1$. 

In $AdS_{4}$ ($d = 3$) there are an infinite set of roots of Eq. (\ref{31}) for $F_{d = 3}(\Delta_{\psi})$ from (\ref{27}) and $F_{d = 3}(\Delta^{\rm conf}_{\sigma}) = - 1/2\pi$ from (\ref{27}), (\ref{30}), however all of them violate unitarity bound.

\subsection{\bf{$O(N)$ symmetric model with composite scalar.}}

\qquad It was shown in \cite{Giombi2} that calculations of Witten diagrams essentially simplify when they include "extremal" vertexes \cite{Hoker} that means fulfillment of "extremal" relation (\ref{3}) $\Delta_{\psi_{1}} = \Delta_{\psi_{2}} + \Delta_{\psi_{3}}$ between conformal dimensions of three interacting fields. This in turn means that conformal operator of the first field is the composite operator which is a product of two other operators, see discussion in the Introduction.

In case of validity of the extremal relation $\Delta_{\phi_{1}} = \Delta_{\phi_{2}} + \Delta_{\phi_{3}}$ exponent $\delta_{23}$ (\ref{12}) is equal to zero, thus coefficient $B$ (\ref{13}) includes $\Gamma(0)$, and first expression for ${\cal {\widetilde M}}^{{\rm 2pt \, bubble}}_{\Delta_{\phi_{1}}|\Delta_{\phi_{2}}\Delta_{\phi_{3}}}\,({\vec x}_{1}, {\vec x}_{2})$ in the RHS of (\ref{24}) diverges as $\Gamma^{2}(0)$. To compensate this divergence it is necessary to apply a regularization proposed in \cite{Giombi2} which is different from the one in (\ref{25}). In \cite{Giombi2} the norm-invariant coupling constant was introduced:

\begin{equation}
\label{34}
g^{* 2} = g^{2} \cdot \frac{B^{2}(\Delta_{\phi_{1}}, \Delta_{\phi_{2}}, \Delta_{\phi_{3}})}{C_{\Delta_{\phi_{1}}} C_{\Delta_{\phi_{2}}} C_{\Delta_{\phi_{3}}}}.
\end{equation}
In the "extremal" case $g^{* 2}$ includes $\Gamma^{2}(0)$. It reduces the same divergence in (\ref{24}). Also, thanks to extremal relation (\ref{3}), practically all Gamma functions in the RHS of (\ref{24}) reduce and three bootstrap equations for conformal correlators of fields $\phi_{i}$ ($i = 1, 2, 3$) take extremely simple form:

\begin{eqnarray}
\label{35}
\frac{C_{\Delta_{\phi_{1}}}}{P_{12}^{\Delta_{\phi_{1}}}} = \frac{g^{* 2} \, C_{\Delta_{\phi_{1}}}}{P_{12}^{\Delta_{\phi_{1}}}},  \qquad  \qquad \,  \, \nonumber
\\ \nonumber
\\
\frac{C_{\Delta_{\phi_{2}}}}{P_{12}^{\Delta_{\phi_{2}}}} = \frac{g^{* 2} \, C_{\Delta_{\phi_{2}}}}{P_{12}^{\Delta_{\phi_{2}}}} \, \cdot \, \frac{F(\Delta_{\phi_{1}})}{F (\Delta_{\phi_{2}})},
\\  \nonumber
\\
\frac{C_{\Delta_{\phi_{3}}}}{P_{12}^{\Delta_{\phi_{3}}}} = \frac{g^{* 2} \, C_{\Delta_{\phi_{3}}}}{P_{12}^{\Delta_{\phi_{3}}}} \, \cdot \, \frac{F(\Delta_{\phi_{1}})}{F (\Delta_{\phi_{3}})}, \nonumber
\end{eqnarray}
$F(\Delta)$ see in (\ref{27}).

Now let us again consider $O(N)$ symmetric model of $N$ fields $\psi_{k}$ with equal conformal dimensions $\Delta_{\psi}$ interacting according to (\ref{29}) with the Hubbard-Stratonovich field $\sigma(Z)$. Since $\sigma(Z) \sim \Sigma_{k}\psi_{k}^{2}(Z)$ it may be considered as composite field which conformal dimension obey extremal relation:

\begin{equation}
\label{36}
\Delta_{\sigma} = 2\,\Delta_{\psi}.
\end{equation}

Thus in this case system (\ref{35}) comes to the following spectral equation:

\begin{equation}
\label{37}
N \, F(\Delta_{\psi}) = F(2 \, \Delta_{\psi}),
\end{equation}
whose positive nontrivial roots for $d = 4$, $N = 1, 2, 3$ are:

\begin{eqnarray}
\label{38}
N = 1: \, \, \, \Delta_{\psi} =  4/3; \, \, \, \,  \, 7/5; \, \, \, \, \quad \qquad \nonumber
\\
N = 2: \,\,\, \Delta_{\psi} = 1.4235 \pm i\,0.1486; \,\,\, \,
\\
N = 3: \,\,\, \Delta_{\psi} = 1.4628 \pm i\,0.1883. \,\,\, \, \,  \nonumber
\end{eqnarray} 

And for $d = 3$ positive roots of (\ref{37}) obeying unitarity bound are:

\begin{eqnarray}
\label{39}
N = 1: \, \, \, \Delta_{\psi} =  0.554; \, \, \, \,  \, 1.238; \, \, \, \, 1.749; \, \, \, 2.248; \quad \qquad \nonumber
\\
N = 2: \,\,\, \Delta_{\psi} = 0.530; \, \, \, 1.093; \,\,\, 1.218; \,  \, \, 1;748; \, \, \, 2.247;
\\
N = 3: \,\,\, \Delta_{\psi} = 0.521; \, \, \, 1.747; \, \, \, 2,245. \,\, \, \, \qquad  \qquad \qquad \nonumber
\end{eqnarray} 

Complex conformal dimensions are not something unusual. As a rule it signals about the nonunitarity of the theory, see e.g. \cite{Hogervorst}. However in the recent paper \cite{Metsaev} it is argued that complexity of conformal dimensions guarantees classical unitarity of the continuous-spin field.

\section{Conclusion}

\qquad Numerical results (\ref{33}), (\ref{38}), (\ref{39}) were obtained under questionable assumptions such as the use of simplified bootstrap equations (\ref{28}) and hypotheses (\ref{30}) or (\ref{36}) regarding the conformal dimension of the Hubbard-Stratonovich field. The meaning of these results is to demonstrate that game worth the candle: proposed AdS/CFT version of the "old" conformal bootstrap may predict values of conformal dimensions. Solution of the exact bootstrap equations (\ref{22}) is the task for future.

Another task for future may be to study the possibility of spontaneous $O(N)$ symmetry breaking in the models of type (\ref{29}). This means that every of fields $\psi_{k}$ in (\ref{29}) must be equipped with its own conformal dimension and asymmetric solutions of self-consistent bootstrap equations (\ref{22}) (or of their simplified version (\ref{28})) must be found. 

Surely the problem of missing equation for the renormalized coupling constant is still there. According to general bootstrap ideology coupling constant may be found from the condition of crossing symmetry of the 4-point amplitudes that in the "old" conformal bootstrap is equivalent to second equation in (\ref{1}).

The simplest case of the interacting bulk scalar fields is considered in the paper. However many questions of modern physics are connected with fermions of spin $1/2$, compare Steven Weinberg's words cited in the beginning of the Introduction. The point is that "flavors" mass hierarchy, which is still a mystery, may be explained in frames of the Randall-Sundram model \cite{Randall} when some natural ("twisted") boundary conditions are imposed on the bulk spinor fields and for certain bulk masses of these fields (see e.g. \cite{Neubert}, \cite{Pomarol}). Thus calculation of fermion bulk masses (that is of conformal dimensions) in frames of the proposed approach of the "old" conformal bootstrap may open the way for solution of the fermion mass hierarchy problem.

\section*{Acknowledgments} Author is grateful to Peter Arseev, Andrei Barvinsky, Ruslan Metsaev, Arkady Tseytlin and Mikhail Vasiliev for stimulating comments and to participants of the seminar in the Theoretical Physics Department of P.N. Lebedev Physical Institute for fruitful discussions.


\begin{thebibliography}{99}
\bibitem{Weinberg}S. Weinberg, {\it{"Model physicist"}}, Interview to Matthew Chalmers, CERN Courier, Oct 13, 2017, http://cerncourier.com/cws/article/cern/70138.
\bibitem{Melon1}I.R. Klebanov and G. Tarnopolsky, {\it{"Uncolored Random Tensors, Melon Diagrams, and the SYK Models"}}, Phys. Rev. {\bf{D 95}} (2017) 046004 [ArXiv:hep-th/1611.08915].
\bibitem{Melon2}I.R. Klebanov, A. Milekhin, F. Popov, and G. Tarnopolsky, {\it{"Spectra of Eigenstates in Fermionic Tensor Quantum Mechanics"}}, Phys. Rev. {\bf{D 97}} (2018) 106023 [ArXiv;hep-th/1802.10263].
\bibitem{Sakharov}A.D. Sakharov, {\it{"Vacuum Quantum Fluctuations In Curved Space And The Theory Of Gravitation"}}, Sov. Phys. Dokl. 12 (1968) 1040 [Dokl. Akad. Nauk Ser. Fiz. 177 (1968) 70]. Reprinted in Gen. Rel. Grav. 32 (2000) 365-367.
\bibitem{old1}A.M. Polyakov, Zh. Exp. Theor. Fiz. {\bf{55}} (1968) 1026 [Sov. Phys. JETP {\bf{28}} (1969) 533].
\bibitem{old2}A.A. Migdal, Zh. Exp. Theor. Fiz. {\bf{55}} (1968) 1964 [Sov. Phys. JETP {\bf{28}} (1969) 1036]; {\it{"Conformal invariance and bootstrap"}}, Phys. Lett. {\bf{37B}} (1971) 386-388; {\it{"Ancient History of CFT"}}, http://migdal100.itp.ac.ru/proceedings-files-MigdalAA-AncientHistoryOfCFT.pdf. 
\bibitem{Parisi}G. Parisi and L. Peliti, Lett. Nuovo Cim. {\bf{2}} (1971) 627; M. d'Eramo, G. Parisi, and L. Peliti, Lett. Nuovo Cim. {\bf{2}} (1971) 878; G. Parisi, {\it{"On self-consistency conditions in conformal covariant field theory"}}, Lett. Nuovo Cim. {\bf{4S2}} (1972) 777-780.
\bibitem{Fradkin}E.S. Fradkin and M.Ya. Palchik, {\it{Conformal quantum field theory in $D$ dimensions}} (Kluwer Academic Publ., Dordrecht, 1996); E.S. Fradkin and M.Ya. Palchik, Phys. Report {\bf{44C}} (1978) 249.
\bibitem{Symanzik}G. Mack and K. Symanzik, {\it{"Currents, stress tensor and generalized unitarity in conformal invariant quantum field theory"}}, Comm. Math. Phys. {\bf{27}} (1972) 247.
\bibitem{Todorov}G. Mack and T. Todorov, {\it{"Conformal-invariant Green functions without ultraviolet divergences"}}, Phys. Rev. {\bf{D 8}} (1973) 1764-1787.
\bibitem{Dobrev1}V.K. Dobrev, V.B. Petkova, S.G. Petrova and I.T. Todorov, {\it{"Dynamical derivation of vacuum operator product expansion in Euclidean conformal quantum field theory"}}, Phys. Rev. {\bf{D 13}} (1976) 887-912.
\bibitem{Dobrev2}V.K. Dobrev, G. Mack, V.B. Petkova, S.G. Petrova and I.T. Todorov, {\it{"Harmonic Analysis on the n-Dimensional Lorenz Group and its Applications to Conformal Quantum Field Theory"}}, Lecture Notes in Physics, No 63, 280 pages (Springer Verlag, Berlin-Heidelberg-New York, 1977).
\bibitem{Grensing}D. Grensing and G. Grensing, {\it{"Critical indices and the conformal invariant bootstrap method"}}, Phys. Rev. {\bf{D 18}} (1978) 2890-2900.
\bibitem{Wilson}K.G. Wilson, {\it{"The Renormalization Group and Critical Phenomena"}}, Nobel Lecture, 8 December 1982, https://www.nobelprize.org/prizes/physics/1982/wilson/lecture/.
\bibitem{Ferrara}S. Ferrara, A.F. Grillo, and R. Gatto, {\it{"Tensor representation of conformal algebra and conformally covariant operator product expansion"}}, Annals Phys. {\bf{76}} (1973) 161-188.
\bibitem{Polyakov}A.M. Polyakov, {\it{"Nonhamiltonian approach to conformal quantum field theory"}}, Zh. Exp. Theor. Fiz. {\bf{66}} (1974) 23-42 [Sov. Phys. JETP {\bf{39}} (1974) 9-18].
\bibitem{Mack}G. Mack, {\it{"Convergence of Operator Product Expansion on the Vacuum in Conformal Invariant Quantum Field Theory"}}, Commun. Math. Phys. {\bf{53}} (1977) 155.
\bibitem{Rychkov}D. Poland, S. Rychkov, and A. Vichi, {\it{"The Conformal Bootstrap: Theory, Numerical Techniques, and Applications"}}, [ArXiv: hep-th/1805.04405].
\bibitem{Alday}L.F. Alday and A. Zhiboedov, {\it{"Conformal Bootstrap With Slightly Broken Higher Spin Symmetry"}}, JHEP 1606 (2016) 091 [ArXiv:hep-th/1506.04659].
\bibitem{Giombi1}S. Giombi, C. Sleight, and M. Taronna, {\it{"Spinning AdS Loop Diagrams: Two Point Functions"}}, JHEP, 1806 (2018) 030 [ArXiv: hep-th/1708.08404].
\bibitem{Giombi2}S. Giombi, V. Kirilin, and E. Perlmutter, {\it{"Double-Trace Deformations of Conformal Correlations"}}, JHEP 1802 (2018) 175 [ArXiv: hep-th/1801.01477].
\bibitem{Mitra}S.S. Gubser and I. Mitra, {\it{Double-trace operators and one loop vacuum energy in AdS/CFT}}, Phys. Rev. {\bf D67} (2003) 064018 [ArXiv: hep-th/0210093].
\bibitem{Hartman}T. Hartman and L. Rastelli, {\it{"Double-trace deformations, mixed boundary conditions and functional determinants in AdS/CFT"}}, JHEP 0801 (2008) 019 [ArXiv: hep-th/0602106].
\bibitem{Diaz}D.E. Diaz and H. Dorn, {\it{"Partition functions and double-trace deformations in AdS/CFT"}}, JHEP 0705 (2007) 046 [ArXiv: hep-th/0702163].
\bibitem{Saharian}A.A. Saharian, {\it{Wightman function and Casimir densities on AdS bulk with application to the Randall-Sundrum braneworld}}, Nucl. Phys. {\bf{B 712}} (2005) 196 [ArXiv:hep-th/0312092].
\bibitem{Alt1}B.L. Altshuler, {\it{"Sakharov's induced gravity on the AdS background. SM scale as inverse mass parameter of Schwinger-DeWitt expansion"}}, Phys. Rev. D {\bf 92} (2015) 065007 [ArXiv: hep-th/1505.07421].
\bibitem{Alt2}B.L. Altshuler, {\it{"Simple way to calculate $UV$-finite one-loop quantum energy in Randall-Sundrum model"}}, Phys. Rev. {\bf{D 95}} (2017) 086001 [ArXiv:hep-th/1701.01541]; {\it{"Scalar field on AdS: quantum one loop "in one line"}}, Report at the Ginzburg Centennial Conference, Lebedev Institute, Moscow, 29 May - 03 June 2017 [ArXiv:hep-th/1706.06286].
\bibitem{Allais}A. Allais, {\it{"Double-trace deformations, holography and the $c$-conjecture"}}, JHEP {\bf{11}} (2010) 040[ArXiv: hep-th/1007.2047].
\bibitem{Diaz2}R. Aros, D.E. Diaz, {\it{"Determinant and Weyl anomaly of Dirac operator: a holographic derivation"}}, [ArXiv:math-ph/1111.1463].
\bibitem{Keldysh}L.V. Keldysh, JETP, {\bf 47} (1964) 1515; Sov. Phys., JETP {\bf 20} (1965) 1018.
\bibitem{Arseev}P.I. Arseev, {\it{"On the nonequilibrium diagram technique: derivation, some features, and applications"}}, Physics-Uspekhi, {\bf 58} (12) (2015).
\bibitem{Barvinsky}A.O. Barvinsky, {\it{"Serendipitous discoveries in nonlocal gravity theory"}}, Phys. Rev. {\bf D 85} (2012) 104018 [ArXiv:hep-th/1112.4340].
\bibitem{Meulen}M. Meulen, J. Smit, {\it{"Classical approximation to quantum cosmological correlations"}}, JCAP (2007) 0711:023 [ArXiv:hep-th/0707.0842].
\bibitem{Akhmedov}E.T. Akhmedov, {\it{"IR divergences and kinetic equation in de Sitter space. Poincare patch: Principal series"}}, JHEP 1201 (2012) 066 [ArXiv:hep-th/1110.2257]; E.T. Akhmedov, F.K. Popov and V.M. Slepukhin, {\it{"Infrared dynamics of the massive $\phi^{4}$ theory on de Sitter space"}}, Phys. Rev. {\bf D 88} (2013) 024021 [ArXiv:hep-th/1303.1068].
\bibitem{Hoker}E. D'Hoker, D.Z. Freedman, S.D. Mathur, A. Matusis, and L. Rastelli, {\it{"Extremal correlators in the AdS/CFT correspondence"}}, Published in {\it{"The Many Faces of the Superworld: Yuri Golfand Memorial Volume"}}, Edited by M.A. Shifman, World Scientific Pub Co Inc. 2000  [ArXiv:hep-th/9908160].
\bibitem{Penedones}J. Penedones, {\it{"Writing CFT correlation functions as AdS scattering amplitudes"}}, JHEP 1103 (2011) 025 [ArXiv:hep-th/1011.1485]. 
\bibitem{Fitz}A.L. Fitzpatrick, J. Kaplan, J. Penedones, S. Raju, B.C. van Rees, {\it{"A Natural Language for AdS/CFT Correlators"}}, JHEP (2011) 2011:95 [ArXiv:hep-th/1107.1499].
\bibitem{Costa}M.S. Costa, V. Goncalves and J. Penedones, {\it{"Spinning AdS Propagators"}}, JHEP 1409 (2014) 064 [ArXiv:hep-th/1404.5625].
\bibitem{Bekaert}X. Bekaert, J. Erdmenger, D. Ponomarev and S. Sleight, {\it{"Towards holographic higher-spin interactions: Four-point functions and higher-spin exchange"}}, JHEP {\bf 03} (2015) 170 [ArXiv:hep-th/1412.0016].
\bibitem{Freedman}D.Z. Freedman, S.D. Mathur, A. Matusis, and L. Rastelli, {\it{"Correlation functions in the $CFT_{d}/AdS_{d+1}$ correspondence"}}, Nucl. Phys. {\it{B 546}} (1999) 96-118 [ArXiv:hep-th/9804058].
\bibitem{Paulos}M.F. Paulos, {\it{"Towards Feynman rules for Mellin amplitudes in AdS/CFT"}}, JHEP (2011) 2011:74 [ArXiv:hep-th/1107.1504].
\bibitem{Symanzik2}K. Symanzik, {\it{On calculations in conformal invariant field theories}}, Lett. Nuovo Cim. {\bf{3}} (1972) 734-738.
\bibitem{Hogervorst}M. Hogervorst, S. Rychkov, B.C. van Rees, {\it{"Unitarity violation at the Wilson-Fisher fixed point in $4 - \epsilon$ dimensions"}}, Phys. Rev. {\bf{D93}} (2015) 125025 [ArXiv:hep-th/1512.00013].
\bibitem{Metsaev}R.R. Metsaev, {\it{"Light-cone continuous-spin field in AdS space"}} [ArXiv:hep-th/1903.10495].
\bibitem{Randall}L. Randall, R. Sundrum, {\it{"A Large Mass Hierarchy from a Small Extra Dimension"}}, Phys.Rev.Lett. {\bf 83}:3370-3373 (1999) [ArXiv: hep-ph/9905221].
\bibitem{Neubert}Y. Grossman and M. Neubert, {\it{"Neutrino masses and mixings in non-factorizable geometry"}}, Phys. Lett. {\bf{B474}} (2000) 361 [ArXiv:hep-ph/9912408].
\bibitem{Pomarol}T. Gherghetta and A. Pomarol, {\it{"Bulk fields and supersymmetry in a slice of AdS"}}, Nucl.Phys. {\bf B586} (2000) 141–162 [ArXiv: hep-ph/0003129].
\end{thebibliography}
\end{document}